\documentclass[twoside,11pt]{article}
\usepackage{extsizes}
\usepackage[super,sort&compress,comma]{natbib} 
\usepackage[version=3]{mhchem}
\usepackage[left=1.5cm, right=1.5cm, top=1.785cm, bottom=2.0cm]{geometry}
\usepackage{balance}
\usepackage{tcolorbox}
\usepackage{etoolbox}
\usepackage{sectsty}
\usepackage{graphicx} 
\usepackage{lastpage}
\usepackage[format=plain,justification=justified,singlelinecheck=false,font={stretch=1.125,small,sf},labelfont=bf,labelsep=space]{caption}
\usepackage{float}
\usepackage{fancyhdr}
\usepackage{fnpos}
\usepackage[english]{babel}
\addto{\captionsenglish}{%
  
}
\usepackage{array}
\usepackage{droidsans}
\usepackage{charter}
\usepackage[T1]{fontenc}
\usepackage[usenames,dvipsnames]{xcolor}
\usepackage{setspace}
\usepackage[compact]{titlesec}
\usepackage{hyperref}

\definecolor{cream}{RGB}{222,217,201}

\usepackage{lmodern}
\usepackage{amsmath, amssymb}
\usepackage{physics}
\usepackage{siunitx}

\newcommand{\epsr}{\epsilon_r}
\newcommand{\epsp}{\epsilon_{\varphi}}
\newcommand{\epsz}{\epsilon_z}
\newcommand{\epsb}{\epsilon_{\text{bulk}}}
\newcommand{\mt}{\tilde{m}}

\newcommand{\qt}{\tilde{q}}
\newcommand{\qtl}{\tilde{q}_l}
\newcommand{\qts}{\tilde{q}_s}

\newcommand{\ind}{\text{ind}}
\newcommand{\ext}{\text{ext}}
\newcommand{\tot}{\text{tot}}
\newcommand{\ing}{\text{in}}
\newcommand{\outg}{\text{out}}
\newcommand{\eff}{\text{eff}}

\newcommand{\td}{\text{2D}}
\newcommand{\tdeg}{\text{2DEG}}
\newcommand{\me}{m_e^*}

\titleformat{\section}          
  {\bfseries\fontsize{12pt}{14pt}\selectfont} 
  {\thesection}                
  {1em}                        
  {}                           
\begin{document}

\begin{flushleft}
{\fontsize{18pt}{20pt}\selectfont\textbf{Electrostatic Screening in Nanotubes: A Tubular Response Function Framework}}
\end{flushleft}
{\fontsize{12pt}{14pt}\selectfont Peter Gispert\textsuperscript{1} and Nikita Kavokine\textsuperscript{*1}}\\\\

\footnotetext[1]{The Quantum Plumbing Lab (LNQ), \'Ecole Polytechnique F\'ed\'erale de Lausanne (EPFL), 1015 Lausanne, Switzerland. \\ $^*$E-mail: nikita.kavokine@epfl.ch.} 

The structure and transport of electrolytes in nanoscale channels are known to be affected by the electronic properties of the confining walls. This influence is particularly pronounced in quasi-one-dimensional nanotubes, where the high surface-to-volume ratio makes the wall the dominant source of electrostatic screening. For instance, ideal metallic tubes suppress long-range Coulomb interactions between ions exponentially. Yet, there exists no generic framework for evaluating electrostatic interactions in tubular confinement.
Here, we introduce tubular response functions -- a generalisation of surface response functions that captures how nanotubes with arbitrary electronic properties screen Coulomb interactions. Using this framework, we evaluate the interaction potential of ions confined in a metallic carbon nanotube, treating its long-range electronic properties exactly within a Luttinger liquid model.  
We demonstrate that the screening characteristic of metallic armchair carbon nanotubes is almost identical to that of an ideal metal, regardless of electron density. We trace the origin of such strong screening to the quantum confinement of electrons around the tube circumference and to the suppression of Friedel oscillations. Our framework opens the way for quantitative descriptions of ionic correlations and charge storage in nanotube-based electrodes, and can be further extended to address confined ion dynamics.

\vspace{1cm}

\section*{Introduction}
Ions are at the heart of many nanofluidic applications, from desalination \cite{poradaReviewScienceTechnology2013} to osmotic power generation \cite{siriaGiantOsmoticEnergy2013} to electrical double-layer supercapacitors \cite{simonMaterialsElectrochemicalCapacitors2008}. These technologies leverage the enormous surface-to-volume ratio of nanoscale channels, exploiting the particularities of confined solid-liquid interfaces, including the surface charge and electric polarisability. Such interfacial effects become particularly pronounced in single-digit nanopores, i.e., channels with a diameter of less than \SI{10}{\nm}, where the surface charge can render the channel impermeable to particular ionic species due to electrostatic repulsion \cite{schochTransportPhenomenaNanofluidics2008, bocquetNanofluidicsBulkInterfaces2010}. Even in the absence of surface charge, the dehydration energy hinders ion entry when the spatial confinement becomes comparable to the ion size. Moreover, the substantial dielectric contrast between water and typical channel materials can lead to a self-energy barrier, reinforcing ion-ion interactions and inhibiting ion entry below a few nanometers of confinement \cite{robinIonFillingOnedimensional2023, kavokineInteractionConfinementElectronic2022, robinModelingEmergentMemory2021, teberTranslocationEnergyIons2005}. This has been identified very early on to have important consequences for ion transport through biological lipid membranes \cite{parsegianEnergyIonCrossing1969}.

In contrast to the aforementioned ion exclusion phenomena, capacitive ion accumulation in nanoporous carbon materials is most efficient in the smallest accessible pores, where the pore diameter matches the size of the ion \cite{largeotRelationIonSize2008, chmiolaAnomalousIncreaseCarbon2006}. This striking experimental observation has been attributed to the so-called \emph{superionic state} \cite{kornyshevDoubleLayerIonicLiquids2007}, promoted by the polarisability of carbon-based nanopores. Kondrat et al. demonstrated a substantial attraction between an ion and its induced image charge in the wall for metallic 2D nanoslits and 1D nanopores \cite{kondratTheorySimulationsIonic2023}. 
Moreover, the image charges in a perfect metal exponentially screen long-range interionic interactions, facilitating the dense packing of like-charge ions \cite{rochesterInterionicInteractionsConducting2013, locheGiantAxialDielectric2019}. Such exponential screening is preserved in nanochannels confined by a Thomas-Fermi metal, a model for an imperfect bulk metal, albeit with an increased screening length \cite{rochesterInterionicInteractionsConducting2013}.

The dielectric constant of water, as a solvent, is strongly modified under nanoconfinement compared to its isotropic value $\epsilon \approx 80$ in the bulk. For 2D slits, Fumagalli et al. have experimentally measured both the out-of-plane \cite{fumagalliAnomalouslyLowDielectric2018} and in-plane \cite{wangInplaneDielectricConstant2025} dielectric constant. They find that the out-of-plane dielectric constant drops with decreasing slit height from its bulk value above $\sim \SI{100}{\nm}$ down to $\epsilon_{\perp} \approx 2$ at a few nanometer height. In contrast, the in-plane dielectric constant rises with decreasing slit height from the bulk value above $\sim \SI{10}{\nm}$ up to $\epsilon_{\parallel} \approx 1000$ at a few nanometer height.
For the nanotube geometry, to our knowledge, no experimental data is available. However, molecular dynamics simulations with SPC/E water in frozen and unpolarisable carbon nanotubes give qualitatively similar results, where the radial dielectric constant drops down to $\epsilon_r \sim 1-3$ and the axial dielectric constant rises up to $\epsilon_z \sim 10^3$ at the radius $R \sim \SI{1}{\nm}$ \cite{locheGiantAxialDielectric2019}.

For two-dimensional nanochannels, the treatment of interionic interactions and electronic screening has recently been generalised to capture arbitrary electronic properties of the confining walls \cite{kavokineInteractionConfinementElectronic2022}. This formalism is based on surface response functions \cite{pitarkeElectronEnergyLoss1997a}, which encode the dielectric response of an arbitrary material across a planar interface. The surface response function can be derived from either microscopic or macroscopic models, and it acts as a reflection coefficient for an evanescent plane-wave potential incident on the interface \cite{kavokineInteractionConfinementElectronic2022}.

Focusing on cylindrical nanotubes, early studies have derived the dielectric function and plasmon dispersion for a free electron gas on a cylindrical shell in the random-phase approximation \cite{linElementaryExcitationsCylindrical1993, longeCollectiveExcitationsMetallic1993}, as well as the radial and axial polarizability of carbon nanotubes \cite{benedictStaticPolarizabilitiesSinglewall1995}. Later, the low-frequency dynamical dielectric function of an arbitrary carbon nanotube (CNT) was analysed \cite{linLowfrequencyElectronicExcitations2000}.
CNTs exhibit complex band structures and electronic properties due to their diverse chiralities \cite{andoTheoryElectronicStates2005, charlierElectronicTransportProperties2007}. This is because a CNT can be viewed as a sheet of graphene rolled into a cylindrical tube, where the orientation of the tube axis on the graphene lattice is defined by two indices $(n,m)$. These indices fully determine the electronic properties of the CNT: Depending on the choice of indices $(n,m)$, the CNT can either be metallic, with an approximately linear band dispersion around the charge-neutrality point, or semiconducting.
For quasi-one-dimensional nanotubes, electron-electron interactions are strongly enhanced by geometrical confinement, promoting long-range correlations and power-law response functions characteristic of one-dimensional systems \cite{giamarchiQuantumPhysicsOne2003}. In the long-wavelength, low-energy limit, these interactions can be treated exactly using bosonization, leading to the Tomonaga-Luttinger description of interacting 1D conductors \cite{haldaneLuttingerLiquidTheory1981, luttingerExactlySolubleModel1963, tomonagaRemarksBlochsMethod1950, mattisExactSolutionManyFermion1965}. This quasi-1D framework is fundamentally different from the Fermi-liquid model, valid in higher dimensions: In a Luttinger liquid, spin and charge degrees of freedom decouple, and fermionic quasi-particles are absent near the Fermi surface, so that the low-energy excitations become collective bosonic modes \cite{giamarchiQuantumPhysicsOne2003, voitOneDimensionalFermiLiquids1995}. 
The Luttinger-liquid framework has been explicitly applied to carbon nanotubes, where strong electron–electron interactions lead to the characteristic power-law suppression of the tunnelling density of states \cite{yaoCarbonNanotubeIntramolecular1999, eggerEffectiveLowEnergyTheory1997, eggerLuttingerLiquidBehavior1999, bockrathLuttingerliquidBehaviourCarbon1999, kaneCoulombInteractionsMesoscopic1997}. 
However, it remains unknown how these intricate electronic properties of CNTs affect interactions between ions confined within the tube. 

In this paper, we generalise surface response functions, developed for planar interfaces, to \emph{tubular response functions} for the cylindrical geometry of nanotubes. We first demonstrate how the tubular response functions emerge from the linear-response properties of a material. We then derive explicit expressions for the tubular response function of various materials, including an insulating, dielectric nanopore, a 1D cylindrical electron gas and a metallic armchair carbon nanotube. For each material, we derive the screened potential of an ion at the centre of the tube, and we deduce the functional form of the decay law at long distances.

\section{Evaluation of the confined ion potential}
\begin{figure}
	\centering
	\includegraphics{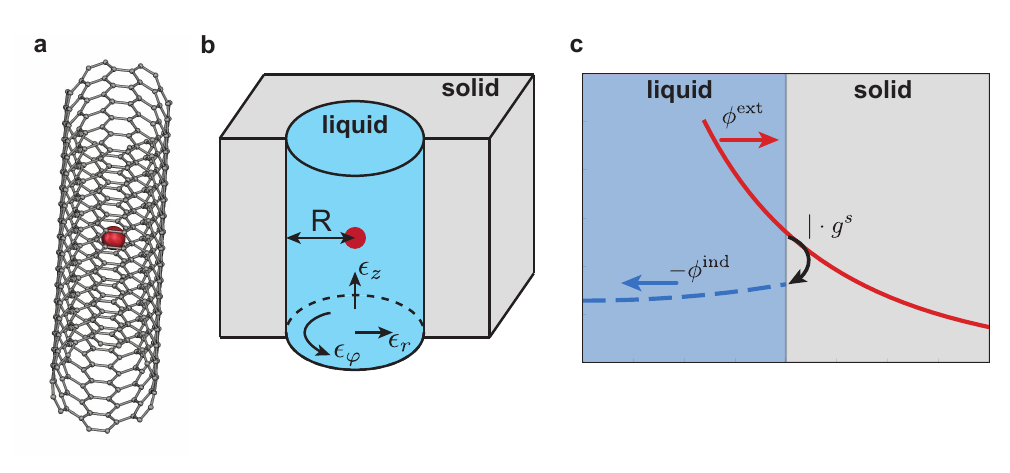}
	\caption{\textbf{a} Schematic illustration of an ion in a (6,6) armchair carbon nanotube. \textbf{b} Schematic of a nanopore of radius $R$ embedded into an infinite dielectric medium. The radial, azimuthal and axial dielectric constants, $\epsr$, $\epsp$ and $\epsz$, are indicated with arrows. \textbf{c} Illustration of the tubular response function $g_s$ as a reflection coefficient for the external potential $\phi^{\ext}$ applied to the solid from the liquid side. The external potential polarises the solid, leading to an induced potential $\phi^{\ind}$ in the liquid. At the solid-liquid interface, the induced potential is $\phi^{\ind}(R) = -g^s \phi^{\ext}(R)$.
	}
	\label{fig1}
\end{figure}

We consider a nanotube of radius $R$ with an ion of charge $+Qe$ placed at its centre (Fig.~\ref{fig1}a). We choose cylindrical coordinates $(\rho, \varphi, z)$, with the $z$-coordinate aligned along the tube. The tube is filled with water, which we treat as an anisotropic dielectric background where $\epsr$, $\epsp$ and $\epsz$ are respectively the radial, azimuthal and axial components of the dielectric tensor $\vb*{\epsilon}$ (Fig.~\ref{fig1}b). Since we consider quasi-1D nanotubes in this paper, we exemplarily set the dielectric constants to $\epsr = \epsp = 2$ and $\epsz = 80$. In contrast, for large tube radii we expect them to converge to the values of a flat interface, $\epsr \approx 2$ and $\epsp = \epsz \approx 80$.

We aim to compute the screening of the Coulomb potential of the ion inside the "liquid" due to the presence of an arbitrary nanotube material, the "solid". Phenomenologically, the ion polarises the nanotube, generating an induced potential that acts back on the nanotube's interior. The derivation can be divided into three steps, as illustrated in Fig.~\ref{fig1}c: (i) Inside the liquid, compute the Coulomb potential of the ion and find the potential at the solid-liquid interface, (ii) inside the solid, compute the dielectric response to the ion potential (decomposed into evanescent cylindrical waves) and find the resulting induced potential at the solid-liquid interface, (iii) compute how the confined liquid screens the potential induced by the solid. Across the solid-liquid interface, the Coulomb potential is continuous. The total potential in the liquid is the sum of the bare potential and the induced potential.

The structure of this problem is generic for any solid-liquid interface geometry, and the formalism for arbitrary electronic properties of the solid has previously been developed for planar surfaces \cite{kavokineInteractionConfinementElectronic2022, coquinotCollectiveModesQuantum2024}. 
For the tubular geometry, we first decompose the Coulomb potential at position $(\rho,\varphi,z)$ of a point charge $Qe$ at position $(\rho',\varphi',z')$ into its cylindrical eigenmodes
\begin{equation}
\label{eq:Coulomb-cylinder}
	\phi(\vb{r}) = \frac{Qe}{(2\pi)^2} \sum_{m \in \mathbb{Z}} \int_{\mathbb{R}} \dd{q} G_{m,q}(\rho,\rho') e^{im(\varphi-\varphi')} e^{iq(z-z')} 
	\qq{with} G_{m,q}(\rho,\rho') = \frac{1}{\epsilon_0 \epsr}I_{\mt}(\qt \rho_<) K_{\mt}(\qt \rho_>),
\end{equation}
where $\mt = \sqrt{\epsp/\epsr} \abs{m}$, $\qt = \sqrt{\epsz/\epsr} \abs{q}$, $\rho_< = \min\{\rho,\rho'\}$, $\rho_> = \max\{\rho,\rho'\}$, and $I_{\nu}$ and $K_{\nu}$ are the modified Bessel functions of order $\nu$.
Here, $G_{m,q}(\rho,\rho')$ is the Fourier transform with respect to $\varphi-\varphi'$ and $z-z'$ of the Coulomb kernel satisfying the Poisson equation $\nabla_{\vb{r}} [\vb*{\epsilon} \cdot \nabla_{\vb{r}} G(\vb{r},\vb{r'})] = - \delta(\vb{r}-\vb{r'})$.
Due to the decomposition in Eq.~\eqref{eq:Coulomb-cylinder}, it is sufficient to study an individual eigenmode $(m,q)$ of the ion potential. 

For step (i), we apply Eq.~\eqref{eq:Coulomb-cylinder} to an ion at the centre of the tube, yielding the external potential inside the liquid
\begin{equation}
\label{eq:bare-ion-potential}
	\phi^{\ext}(\rho,\varphi,z)
	= \frac{Qe}{(2\pi)^2} \sum_{m \in \mathbb{Z}} e^{im\varphi} \int_{\mathbb{R}} \dd{q} e^{iqz} \, G_{m,q}(\rho, 0) 
	= \frac{Qe}{4\pi \epsilon_0 \sqrt{\epsr}} \frac{1}{\sqrt{\epsz \rho^2 + \epsr z^2}} .
\end{equation}
It is worth noting that the potential strength along the tube axis ($\rho = 0$) is determined by $\epsr$. Under \aa ngstr\"om-scale confinement, $\epsr$ can be substantially reduced compared to the bulk, leading to an increased Coulomb interaction strength.
The eigenmodes of the external potential at the solid-liquid interface are 
\begin{equation}
	\phi^{\ext}_{m, q}(R) = Qe \, G_{m,q}(R,0) \qq{with} G_{m,q}(R,0) = \frac{1}{\epsilon_0 \epsr}K_0(\qt R) \delta_{m,0} .
\end{equation}
The external potential inside the solid then reads
\begin{equation}
\label{eq:impinging-potential}
	\phi^{\ext}_{m, q}(\rho) = W^{\outg}_{m,q}(\rho) \cdot \phi^{\ext}_{m,q}(R) \qq{with} W^{\outg}_{m,q}(\rho) = \frac{K_{\mt}(\qt\rho)}{K_{\mt}(\qt R)} .
\end{equation}
For step (ii), and in analogy with surface response functions for planar interfaces \cite{pitarkeElectronEnergyLoss1997a, kavokineInteractionConfinementElectronic2022}, we encode the dielectric response of the nanotube in a \emph{tubular response function} $g_{m,q} = -\phi^{\ind}_{m, q}(R) / \phi^{\ext}_{m, q}(R)$. Phenomenologically, the tubular response function acts as a reflection coefficient of the external ion potential impinging on the solid (Fig.~\ref{fig1}c).
For step (iii), the potential applied to the liquid by the (polarised) solid is, analogous to Eq.~\eqref{eq:impinging-potential},
\begin{equation}
\label{eq:reflected-potential}
	\phi^{\ind}_{m, q}(\rho) = W^{\ing}_{m,q}(\rho) \cdot \phi^{\ind}_{m,q}(R) \qq{with} W^{\ing}_{m,q}(\rho) = \frac{I_{\mt}(\qt\rho)}{I_{\mt}(\qt R)}.
\end{equation}
The functions $W^{\ing}_{m,q}(\rho)$ and $W^{\outg}_{m,q}(\rho)$ respectively describe the radial decay of an evanescent cylindrical wave when going towards or away from the tube axis. 
The induced potential inside the liquid thus becomes
\begin{align}
\label{eq:induced-ion-potential}
	\phi^{\ind}(\rho,\varphi,z)
	&= -\frac{2}{(2\pi)^2} \sum_{m \in \mathbb{Z}} e^{im\varphi} \int_0^{\infty} \dd{q} \cos(qz) \qty[ W^{\ing}_{m,q}(\rho) \, g_{m,q} \, \phi^{\ext}_{m,q}(R)] \\
	&= -\frac{2Qe}{\epsilon_0 \epsr (2\pi)^2} \int_0^{\infty} \dd{q} \cos(qz) \frac{I_0(\qt \rho)}{I_0(\qt R)}g_{0,q} K_0(\qt R) .
\end{align}
Conceptually, the term in brackets in Eq.~\eqref{eq:induced-ion-potential} is the product of three factors, which represent the steps (i) to (iii) from right to left. The total potential inside the liquid is $\phi^{\tot}(\rho,\varphi,z) = \phi^{\ext}(\rho,\varphi,z) + \phi^{\ind}(\rho,\varphi,z)$. This formalism is not limited to the description of a single static ion in the centre of the tube. The evanescent electrical potential created by any time-dependent charge distribution inside the tube can be decomposed into cylindrical modes $\phi^{\ext}_{m, q, \omega} \, W^{\outg,s}_{m,q}(\rho) \, e^{im\varphi} e^{iqz} e^{-i \omega t}$ at the interface, and the induced potential is determined by Eq.~\eqref{eq:induced-ion-potential} with a frequency-dependent $g_{m,q,\omega}$. Moreover, the framework of tubular response functions also applies to charges inside the solid. In this case, the charges polarise the liquid, as encoded in the tubular response function of the liquid, $g_{m,q,\omega}^l$. The induced potential inside the solid is then determined by Eq.~\eqref{eq:induced-ion-potential} under the exchange $g \to g^l$ and $W^{\ing} \to W^{\outg}$.
In the formalism presented so far, the electrostatic response properties of the nanotube are entirely encapsulated into the tubular response function. We now proceed with a derivation of the tubular response function from linear response theory.

\section{Tubular response functions}
\subsection{Dielectric medium}
As a first example, we derive the tubular response function of an infinite, purely dielectric medium surrounding a cylindrical pore of radius $R$ (Fig.~\ref{fig1}b). For generality, the liquid and solid may have anisotropic dielectric constants. 
In each region, the electric potential must solve the Poisson equation, where the only charges are the ion in the tube and the induced surface charge at the solid-liquid interface due to the dielectric contrast. Decomposed into cylindrical eigenmodes, the potential must therefore have the form
\begin{equation}
\label{eq:potential-dielectric-interface}
	\phi^{\tot}_{m,q}(\rho)
	= \begin{cases}
		\phi^{\ext}_{m,q} W^{\outg}_{m,q}(\rho) + \phi^{\ind}_{m,q} W^{\ing}_{m,q}(\rho) & \text{for } \rho \leq R \\
		(\phi^{\ext}_{m,q} + \phi^{\ind}_{m,q}) \, W^{\outg}_{m,q}(\rho) & \text{for } \rho>R
	\end{cases} .
\end{equation}
In Eq.~\eqref{eq:potential-dielectric-interface} we have already enforced the continuity of the potential across the interface as well as its asymptotic behaviour for $\rho\to\infty$. For readability, we leave it implicit that the values of $\mt$ and $\qt$ inside the weight-functions $W^{\ing/\outg}_{m,q}$ are generally different for the two material regions. The induced potential amplitude $\phi^{\ind}_{m,q}$ is determined by the continuity of the radial component of the displacement field $D_r = -\epsilon_0\epsr \pdv{\rho} \phi^{\tot}_{m,q}(\rho)$, where $\epsr$ is the radial dielectric constant for the respective material. The general solution is straightforward to obtain, but it is a rather lengthy expression in terms of modified Bessel functions. Here, we focus on an ion at the centre of the tube, which is solely determined by the angular mode $m=0$. The resulting tubular response function is
\begin{equation}
\label{eq:gs-dielectric}
	g_{0,q} = -\frac{\phi^{\ind}_{0,q}}{\phi^{\ext}_{0,q}}
	= -\frac{I_{0}(\qtl R)\,\qty[ \kappa_l \, K_{\nu_l}(\qtl R)\, K_{0}(\qts R) - \kappa_s\, K_{0}(\qtl R)\, K_{\nu_s}(\qts R)]}
	{K_{0}(\qtl R)\,\qty[ \kappa_l \, I_{\nu_l}(\qtl R)\, K_{0}(\qts R) + \kappa_s\, I_{0}(\qtl R)\, K_{\nu_s}(\qts R) ]}
\end{equation}
with $l$ and $s$ denoting "liquid" and "solid", and $\qt = \sqrt{\epsz/\epsr}\abs{q}$, $\kappa = \sqrt{\epsz\epsr}$ as well as $\nu = \sqrt{\epsp/\epsr}$ for the respective materials. 
In the limit $R \to \infty$, Eq.~\eqref{eq:gs-dielectric} reduces to
\begin{equation}
\label{eq:gs-dielectric-flat}
	g_{0,q} = \frac{\sqrt{\epsr^s \epsz^s} - \sqrt{\epsr^l \epsz^l}}{\sqrt{\epsr^s \epsz^s} + \sqrt{\epsr^l \epsz^l}} ,
\end{equation}
which is independent of the wavevector $q$ and coincides with the surface response function of a planar interface \cite{kavokineInteractionConfinementElectronic2022} for $\epsr \to \epsilon_{\perp}$ and $\epsz \to \epsilon_{\parallel}$. However, for a finite radius $R$, the tubular response function of a homogeneous dielectric nanopore remains wavevector-dependent. This means that, in contrast to a planar interface, the induced potential is generally not proportional to the external potential in real space.

\subsection{Derivation from linear response theory}
Microscopically, the dielectric response of a solid-state material stems from the dynamic rearrangement of charges -- electrons and nuclei. In linear response theory, this can be characterised by the charge density response function $\chi(\vb{r},\vb{r'},t,t')$, which, given an external potential $\phi^{\ext}(\vb{r'},t')$, determines the induced charge density
\begin{equation}
	n^{\ind}(\vb{r},t) = \int \dd{t'} \int \dd{\vb{r}'} \, \chi(\vb{r},\vb{r'},t,t') \phi^{\ext}(\vb{r'},t') .
\end{equation}
For a nanotube system with translation-invariance along the tube axis, we define the Fourier transform
\begin{equation}
\label{eq:FT-chi}
	\chi_{m,q,\omega}(\rho,\rho') = \int_{\mathbb{R}} \dd{(t-t')} e^{i \omega (t-t')} \int_{0}^{2\pi} \dd{(\varphi-\varphi')} e^{-i m (\varphi-\varphi')} \int_{\mathbb{R}} \dd{(z-z')} e^{-i q (z-z')} \chi(\vb{r},\vb{r'},t,t') .
\end{equation}
This yields
\begin{equation}
	n^{\ind}_{m,q,\omega}(\rho_s) = \int_S \dd{\rho_s'} \rho_s' \, \chi_{m,q,\omega}(\rho_s,\rho_s') \, W^{\outg,s}_{m,q}(\rho_s') \, \phi^{\ext}_{m,q,\omega}(R) ,
\end{equation}
where we explicitly denote the integration over the region $S$ of the solid, and we use $\phi^{\ext}_{m,q,\omega}(\rho_s') = \phi^{\ext}_{m,q,\omega}(R) \, W^{\outg,s}_{m,q}(\rho_s')$.
The induced charge creates an induced potential, which at the interface reads
\begin{equation}
	\phi^{\ind}_{m,q,\omega}(R) = \int_S \dd{\rho_s} \rho_s \,  G_{m,q}^s(R,\rho_s) \, n^{\ind}_{m,q,\omega}(\rho_s),
\end{equation}
where $G_{m,q}^s$ denotes the Coulomb kernel inside the solid.
The resulting expression for the tubular response function reads
\begin{equation}
\label{eq:tubular-response-gs}
	g^s_{m,q,\omega} = -\frac{\phi^{\ind}_{m,q,\omega}(R)}{\phi^{\ext}_{m,q,\omega}(R)}
	= -\int_S \dd{\rho_s} \rho_s \, \int_S \dd{\rho_s'} \rho_s' \, G_{m,q}^s(R,\rho_s) \, \chi^s_{m,q,\omega}(\rho_s,\rho'_s) \, W^{\outg,s}_{m,q}(\rho_s') .
\end{equation}
Reading the formula from right to left, the external potential propagates from the interface into the solid up to radius $\rho_s'$, which, through the density response function $\chi$, induces a charge inside the solid at radius $\rho_s$, creating an induced potential at the interface.
For completeness, the same reasoning leads to the tubular response function of the liquid, which can be obtained by exchanging the roles of the solid and the liquid:
\begin{equation}
\label{eq:tubular-response-gl}
	g^l_{m,q,\omega}
	= -\int_L \dd{\rho_l} \rho_l \, \int_L \dd{\rho_l'} \rho_l' \, G_{m,q}^l(R,\rho_l) \, \chi^l_{m,q,\omega}(\rho_l,\rho'_l) \, W^{\ing,l}_{m,q}(\rho_l') .
\end{equation}
In the derivation of the tubular response functions, Eq.~\eqref{eq:tubular-response-gs} and \eqref{eq:tubular-response-gl}, we have formally separated the (possibly anisotropic) dielectric background, which enters in the Coulomb kernel $G_{m,q}$, and the dynamic rearrangement of free charges, characterized by the density response function $\chi$. A complete description of the interface then requires a self-consistent treatment of the induced potential due to both the dielectric contrast and the polarisability of mobile charges.
We now derive $\chi$ from a quantum-mechanical perspective, which will finally enable us to compute the tubular response function of realistic nanotube systems.

\subsection{Density response function of tubular systems}
\label{ssec:chi}
For a non-interacting electronic system, the density response function can be obtained from the single-particle eigenenergies $E_{\lambda}$ and eigenfunctions $\psi_{\lambda}(\vb{r})$ as\cite{rammerQuantumFieldTheory2007}
\begin{equation}
\label{eq:chi-bare-definition}
	\chi^0(\vb{r}, \vb{r}', \omega) = \sum_{\lambda, \lambda'} \frac{n_F(E_{\lambda}) - n_F(E_{\lambda'})}{E_{\lambda} - E_{\lambda'} + \hbar \omega + i\Gamma} \mel{\lambda}{n(\vb{r})}{\lambda'} \mel{\lambda'}{n(\vb{r'})}{\lambda},
\end{equation}
where $n_F$ is the Fermi-Dirac distribution and $\Gamma$ is a phenomenological broadening, accounting for the finite quasi-particle lifetime due to scattering processes. If electron-electron interactions are taken into account at the mean-field level -- within the so-called random phase approximation (RPA) -- the interacting response function can be obtained in terms of the non-interacting one through an integral Dyson equation:  
\begin{equation}
\label{eq:Dyson-chi}
	\chi^{s}_{m,q,\omega}(\rho,\rho') = \chi^{s,0}_{m,q,\omega}(\rho,\rho') + \int_S \dd{\rho_1} \rho_1 \, \int_S \dd{\rho_2} \rho_2 \, \chi^{s,0}_{m,q,\omega}(\rho,\rho_2) \, G^s_{m,q}(\rho_2,\rho_1) \, \chi^s_{m,q,\omega}(\rho_1,\rho') .
\end{equation}
In general, Eq.~\eqref{eq:Dyson-chi} needs to be solved numerically. However, for a nanotube wall with vanishing radial thickness, Eq.~\eqref{eq:Dyson-chi} yields
\begin{equation}
\label{eq:Dyson-chi-2D}
	\chi^s_{m,q,\omega}(R,R) = \frac{\chi^{s,0}_{m,q,\omega}(R,R)}{1 - G_{m,q}^{\eff}(R,R) \,\chi^{s,0}_{m,q,\omega}(R,R)} .
\end{equation}
To account for the screening of the electron-electron interactions by the liquid, we use $G_{m,q}^{\eff}(R,R) = G_{m,q}^l(R,R)$. Additional dielectric screening, for instance due to high-energy excitations in the solid, can be incorporated by renormalising the dielectric constants entering $G_{m,q}^{\eff}(R,R)$.
Ultimately, the non-interacting tubular response function is given by
\begin{equation}
\label{eq:tubular-response-gs-2D}
	g^{s,0}_{m,q,\omega}
	= - G_{m,q}^{\eff}(R,R) \, \chi^{s,0}_{m,q,\omega}(R,R) ,
\end{equation}
and the Dyson equation, Eq.~\eqref{eq:Dyson-chi-2D}, translates to
\begin{equation}
\label{eq:RPA-tubular-response}
	g_{m,q,\omega}^s = \frac{g^{s,0}_{m,q,\omega}}{1 + g^{s,0}_{m,q,\omega}} .
\end{equation}

\section{Screened potential for model nanotubes}
We now proceed with the computation of tubular response functions for different model materials in the static limit $\omega \to 0$. We apply the results to an ion at the centre of the nanotube and determine the asymptotic functional form of the potential along the cylinder axis. Based on the equations \eqref{eq:bare-ion-potential} and \eqref{eq:induced-ion-potential}, the screened potential inside the nanotube is
\begin{equation}
	\label{eq:total-ion-potential}
	\phi^{\tot}(\rho,\varphi,z)
	= \frac{2Qe}{\epsilon_0 \epsr (2\pi)^2} \int_0^{\infty} \dd{q} \cos(qz) F(q,\rho) \qq{with} F(q,\rho) = \qty[ K_0(\qt \rho) - \frac{I_0(\qt \rho)}{I_0(\qt R)}g_{0,q} K_0(\qt R) ].
\end{equation}
The Bessel function $K_0(z)$ is logarithmically divergent for $z \to 0$, so we regularise $\phi^{\tot}$ at the tube axis by a small value $\rho>0$.
The long-range behaviour of the total potential is determined by the asymptotic behaviour of $F(q,\rho)$ in the limit $q\to 0$. For $q \to 0$, and within RPA, the bare density response function of any metallic nanotube converges to minus the density of states, $-g(E_F)$, at the Fermi level $E_F$. The asymptotic expansions of the Bessel functions $K_0(x)$ and $I_0(x)$ for $x \to 0$ are
\begin{equation}
\label{eq:Bessel-asymptotic-q0}
	K_0(x) = -\ln(x) + \ln(2)-\gamma_E + \mathcal{O}(x^2) \qq{and} I_0(x) = 1 + x^2/4 + \mathcal{O}(x^4),
\end{equation}
where $\gamma_E$ denotes the Euler constant.
Thus, for a finite density of states at the Fermi level, the RPA-renormalised tubular response function, Eq.~\eqref{eq:RPA-tubular-response} converges to 1 for $q \to 0$. In this case, the potential decays faster than $1/|z|$ as $z \to \infty$.
If $g_{0,q} < 1$ for $q \to 0$, then the asymptotic decay law remains $1/\abs{z}$, and the amplitude of the potential is modified by the factor $(1 - g_{0,0})$. In this case, the nanotube asymptotically acts as a dielectric medium.

In the following, we derive explicit formulas for the tubular response function of various systems and compute the resulting screened ion Coulomb potential.
Fig.~\ref{fig2} shows two typical band structures for quasi-1D systems, namely the quasi-1D electron gas and a $(6,6)$ metallic CNT. The resulting tubular response functions and screened ion potentials are depicted in Fig.~\ref{fig3}.

\subsection{Homogeneous dielectric medium}
The tubular response function $g_{0,q}$ of an infinite, homogeneous, dielectric nanopore from Eq.~\eqref{eq:gs-dielectric} is depicted in Fig.~\ref{fig3}a for $\epsilon^s_{r,\varphi,z} = 4$. In the long-wavelength limit $q \to 0$, the response function converges to $g_{0,0} = 1 - \epsr^l / \epsr^s < 1$. Hence, the long-range amplitude of the bare Coulomb potential is modified by the factor $\epsr^l / \epsr^s$. Concretely, for $\epsr^l < \epsr^s$, the dielectric nanopore provides additional potential screening at long distances, whereas for $\epsr^l > \epsr^s$, the nanopore amplifies the Coulomb potential.
The latter effect is reminiscent of \emph{interaction confinement} \cite{kavokineInteractionConfinementElectronic2022}. This phenomenon was initially identified in nanopores with a low dielectric constant filled with a liquid with a high isotropic dielectric constant. In this case, the electric field concentrates within the material with a high dielectric constant, leading to enhanced quasi-1D Coulomb interactions. Accounting for the substantial anisotropy of the dielectric constant of water in \aa ngstr\"om-scale nanotubes, Coulomb interactions are again enhanced compared to bulk water (Fig.\ref{fig3}b,c). Here, however, this is an immediate consequence of the anisotropic dielectric background since, according to Eq.~\eqref{eq:bare-ion-potential}, the Coulomb potential along the cylinder axis is determined by $\epsr < \epsz$. For $\epsr^l > \epsr^s$, the tube walls provide additional interaction enhancement at long distances.

\subsection{Ideal metal}
An ideal metal can be described as a dielectric in the limit of an infinite dielectric constant, i.e. $\epsr^s,\epsp^s,\epsz^s \to \infty$. The tubular response function of an infinite dielectric medium was already derived in Eq.~\eqref{eq:gs-dielectric}. The limit yields $g_{m,q} = 1$, which means that the external potential is fully reflected at all wavelengths. The asymptotic screening law for $\abs{z} \to \infty$ is exponential.
For the case of an isotropic dielectric liquid, the screened Coulomb potential in a perfect metallic nanotube has previously been derived \cite{rochesterInterionicInteractionsConducting2013}. For the anistropic case, we have to consider $\qt = \sqrt{\epsz/\epsr}\, q$ in Eq.~\eqref{eq:total-ion-potential}. Rescaling the integration variable $q$ and the coordinate $z$ in Eq.~\eqref{eq:total-ion-potential}, we recover the formula for the isotropic case. Thus, the existing result can be generalised to the anisotropic case without explicit calculations. We obtain
\begin{equation}
\label{eq:screened-potential-metal}
	\phi^{\tot}(\rho =0,\varphi,z)
	= \frac{Qe}{2\pi \epsilon_0 \sqrt{\epsr \epsz} R} \sum_{n=1}^{\infty} \frac{e^{-\frac{\abs{z}}{R} \sqrt{\frac{\epsr}{\epsz}} k_n}}{k_n \qty[J_1(k_n)]^2 },
\end{equation}
where $k_n$ is the $n$-th zero of the zero-order Bessel function $J_0$, and $J_1$ denotes the first-order Bessel function. Eq.~\eqref{eq:screened-potential-metal} explicitly demonstrates the exponential decay, and at long distances $\abs{z}$ the first term with $k_1 \approx 2.40$ dominates.
Remarkably, the anisotropy of the dielectric constant of water in \aa ngstr\"om-scale nanotubes enhances the Coulomb potential so strongly that, for our choice of parameters, the metallic tube walls suppress the Coulomb interaction below the bulk strength only for distances $\abs{z}$ larger than $\sim \SI{6}{\nm}$ (Fig.\ref{fig3}b,c).

\subsection{Cylindrical quasi-1D electron gas}
\begin{figure}
	\centering
	\includegraphics{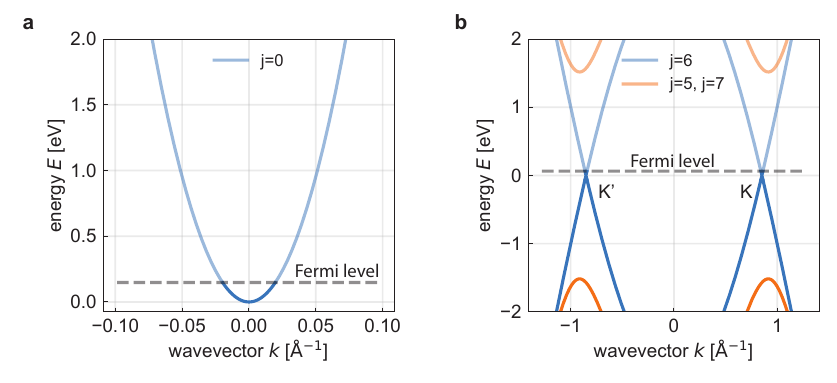}
	\caption{ \textbf{a} Band structure of a quasi-1D electron gas with effective mass $m_e^* = 0.01 m_e$ on an infinite cylinder with radius $R=\SI{4}{\angstrom}$. \textbf{b} Band structure of a (6,6) armchair CNT with radius $R \approx \SI{4}{\angstrom}$. \textbf{a,b} The electron density is $n=\SI{5e12}{cm^{-2}}$.
	}
	\label{fig2}
\end{figure}

As a first quantum-mechanical model, we consider a quasi-1D electron gas (1DEG) confined to the surface of a nanotube with radius $R$. We choose the effective mass $\me = 0.01\, m_e$, where $m_e$ is the bare electron mass. The charge density response function of this system has previously been reported \cite{satoExoticBehaviorDielectric1993}. However, for completeness, we restate the main findings.  
Due to the periodic boundary conditions for the wavefunction around the tube circumference, the momentum in this direction is quantised as $q_j = j / R$ with $j \in \mathbb{Z}$. Thus, the band structure of the electron gas has several bands, indexed by $j$, with energy $E_{j,k} = \frac{\hbar^2}{2 \me} \qty(\frac{j^2}{R^2} + k^2)$, where $k$ is the wavevector along the nanotube axis (Fig.~\ref{fig2}a). The eigenstates $\lambda$ in Eq.~\eqref{eq:chi-bare-definition} can accordingly be labelled as $\lambda = (k,j,\sigma)$, with $\sigma$ denoting spin. Since the eigenfunctions of the electron gas are orthonormal, the non-interacting density response function becomes
\begin{equation}
	\label{eq:chi-bare-1DEG}
	\chi^0_{m, q, \omega} = \frac{g_s e^2}{2\pi} \sum_{j \in \mathbb{Z}} \int_{\mathbb{R}} \dd{k} \frac{n_F(E_{k,j}) - n_F(E_{k+q,j+m})}{E_{k,j} - E_{k+q,j+m} + \hbar \omega + i\Gamma}
\end{equation}
with spin-degeneracy $g_s = 2$.

\begin{figure}
	\centering
	\includegraphics{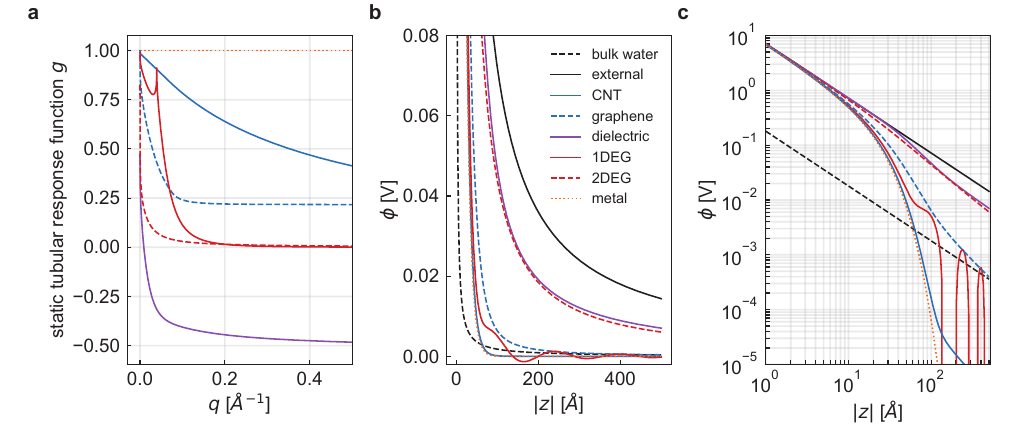}
	\caption{\textbf{Tubular response function and screened ion potential}.  \textbf{a} Tubular response function $g_{0,q}$ for various nanotube systems, defined in the main text, with radius $R=\SI{4}{\angstrom}$. The dielectric medium is isotropic with dielectric constants $\epsilon^s_{r,\varphi,z} = 4$. The perfect metal has, by definition, $g_{0,q}=1$. All other models are based on the same electron density $n=\SI{5e12}{cm^{-2}}$. The armchair CNT has the chiral indices $(6,6)$ and the radius $R \approx \SI{4.07}{\angstrom}$. The effective electron mass of the 1D and 2D free electron gas is $m_e^* = 0.01 m_e$. \textbf{b} Bare and screened Coulomb potential of an ion at the centre of various nanotubes. The potential is computed along the tube axis with the same parameters as for panel \textbf{a}. 
	\textbf{c} Same data as panel \textbf{b}, but in double-logarithmic scale.
	}
	\label{fig3}
\end{figure}

At room temperature, 
the thermal energy $k_B T \approx \SI{25}{\meV}$ is much smaller than the Fermi level $E_F \approx \SI{150}{\meV}$ and the subband spacing. 
We can thus treat the system as if it was at zero temperature, so that the Fermi-Dirac distribution becomes a step function, $n_F(E) = \theta(E_F - E)$, and the integral can be computed analytically. For vanishing broadening $\Gamma=0$, the result is
\begin{equation}
	\chi^0_{m, q, \omega} = \frac{e^2}{\pi} \frac{\me}{\hbar^2 q} \sum_{j} \ln\abs{\frac{[d + 2qk_F(j)][d - 2q(k_F(j)+q)]}{[d - 2qk_F(j)][d + 2q(k_F(j) - q)]}} \qq{with} d = \frac{2jm + m^2}{R^2} + q^2 - \frac{2\me}{\hbar q} \omega,
\end{equation}
where the sum runs over all partially occupied bands, and $k_F(j) = \sqrt{\frac{2\me}{\hbar^2} E_F - (j/R)^2}$ denotes the Fermi wavevector corresponding to band $j$. At the electron density under consideration, only the lowest subband contributes. The resulting tubular response function within RPA, $g_{0,q,\omega=0}$ (see Eqs. \eqref{eq:Dyson-chi-2D}-\eqref{eq:RPA-tubular-response}), is depicted in Fig.~\ref{fig3}a.
In the long-wavelength limit $q \to 0$, the bare density response function $\chi^0_{0,q,\omega=0}$ converges to minus the finite density of states at the Fermi level $g(E_F) = \frac{2e^2}{\pi} \sum_j \frac{\me}{\hbar^2 k_F(j)}$, and therefore the tubular response function converges to 1 (Fig.~\ref{fig3}a). At $q = 2k_F$, the tubular response function features a cusp, originating from interbranch transitions from $-k_F$ to $+k_F$. This gives rise to Friedel oscillations in the screened potential (Fig.~\ref{fig3}b,c).

\subsection{Metallic armchair carbon nanotube: Luttinger framework}

Many nanofluidic systems are based on carbon nanotubes (CNTs) \cite{liCarbonNanotubeNanofluidics2025}. 
We therefore apply the framework of tubular response functions to the electronic screening of ionic interactions inside CNTs, starting from a quantum-level description of the CNT electronic properties. 

To keep the treatment analytical, we start from a nearest-neighbour tight-binding model of the CNT band structure. For details of the model, we refer to the literature \cite{saitoPhysicalPropertiesCarbon2012, andoTheoryElectronicStates2005, charlierElectronicTransportProperties2007}. Here, we focus on so-called armchair CNTs with chiral indices $m=n$. These nanotubes are metallic and at low energies they exhibit the characteristic linear band dispersion in the vicinity of the Dirac points $K$ and $K'$ (Fig.~\ref{fig2}b). As in the quasi-1D electron gas, periodic boundary conditions around the tube circumference give rise to several energy bands. However, for a quasi-1D armchair CNT, the bottom of the second-lowest band has energy $E \approx \SI{10}{\eV}/n$ and thus lies more than $\sim\SI{0.6}{\eV}$ above the charge-neutrality point for nanotubes with radii $R < \SI{1}{\nm}$, corresponding to $n < 15$. We therefore consider only the lowest band, whose dispersion is well approximated by the Dirac fermion dispersion $E_k = \hbar v_F k$. Here, $k$ is measured from the respective Dirac point, and we use $v_F = \frac{3\gamma_0 a_{CC}}{2\hbar} \sim \SI{1e6}{m/s}$, derived from the tight-binding parameter $\gamma_0 = \SI{3.033}{\eV}$ and the carbon-carbon distance $a_{CC} = \SI{1.42}{\angstrom}$ \cite{saitoPhysicalPropertiesCarbon2012}. 

The response properties of an interacting Dirac fermion system in one dimension can be determined exactly in the long-wavelength limit ($q\to 0$) following the Luttinger liquid framework. This problem has been extensively addressed in the literature\cite{luttingerExactlySolubleModel1963, tomonagaRemarksBlochsMethod1950, mattisExactSolutionManyFermion1965, giamarchiQuantumPhysicsOne2003}, and here we outline only the main ideas. In the long-wavelength limit, the Luttinger framework yields an exact transformation that maps the interacting fermionic Hamiltonian onto a non-interacting bosonic Hamiltonian. The bosonic Hamiltonian splits into two parts: $H = H_{\rho} + H_{\sigma}$, corresponding to charge $(\rho)$ and spin ($\sigma$) degrees of freedom. Here, we are interested in the charge excitations, which are governed by \cite{giamarchiQuantumPhysicsOne2003}
\begin{equation}
\label{eq:Luttinger-charge-Hamiltonian}
	H_{\rho} = \sum_q \frac{u(q)}{2\pi L} \qty[ K(q) \pi^2 \Pi^*(q) \Pi(q) + \frac{1}{K(q)} q^2 \phi^*(q) \phi(q)],
\end{equation}
The field $\phi$ is related to the charge density via $\rho(x) = -\pi^{-1} \nabla \phi(x)$, and $\Pi(x)$ is its conjugate momentum. If the fermions are non-interacting, the Fermi velocity reduces to the bare value $u(q) = v_F$, and the Luttinger parameter is $K(q) = 1$. Now, introducing a long-range Coulomb interaction in the initial fermionic Hamiltonian amounts simply to renormalising the parameters of the bosonic one. Accounting for the spin degeneracy $g_s=2$ and the valley degeneracy $g_v = 2$ in the case of CNTs, one obtains\cite{eggerEffectiveLowEnergyTheory1997, voitOneDimensionalFermiLiquids1995, giamarchiQuantumPhysicsOne2003}
\begin{equation}
\label{eq:Luttinger-uK}
	u(q) = v_F \sqrt{1 + \frac{g_s g_v G_{0,q}(R,R)}{\pi v_F}} \qq{and} K(q) = \frac{v_F}{u(q)} .
\end{equation}
The resulting density response function is \cite{voitOneDimensionalFermiLiquids1995}
\begin{equation}
\label{eq:chi-Luttinger}
	\chi_{q,\omega} = \frac{g_s g_v e^2}{\pi \hbar v_F} \frac{(v_F q)^2}{(\hbar \omega)^2 - (u(q) \, q)^2} \xrightarrow{\omega \to 0} - \frac{g_s g_v e^2}{\pi \hbar v_F} \frac{1}{1 + g_s g_v G_{0,q}(R,R)/(\pi v_F)}.
\end{equation}
At this point, we remark that the Hamiltonian in Eq.~\eqref{eq:Luttinger-charge-Hamiltonian} is quadratic in the electronic density: It therefore describes Gaussian charge fluctuations. In section \ref{ssec:chi} we used the random phase approximation (RPA) for treating electron-electron interactions, and RPA is precisely a Gaussian approximation for charge fluctuations. This implies that the seemingly approximate RPA treatment is in fact exact for the density response function of Dirac fermions in 1D in the long-wavelength limit $q\to 0$ \cite{giamarchiQuantumPhysicsOne2003}. 

Going back to Eq.~\eqref{eq:chi-bare-definition} for the non-interacting response function, the matrix elements are in general non-trivial due to the sublattice structure of CNTs. However, in armchair CNTs, the matrix element equals 1 if $k$ and $k+q$ are on the same branch, and zero otherwise. Hence, taking into account both valleys in the Brillouin zone, the bare density response function reads \cite{thakurDynamicalPolarizabilityScreening2017}
\begin{equation}
\label{eq:chi-bare-CNT}
	\chi^0_{0,q,\omega} = \frac{g_s g_v e^2}{\pi \hbar} \frac{v_F q^2}{\hbar^2 (\omega + i\Gamma)^2 - (v_F q)^2} \xrightarrow{\omega,\Gamma \to 0} -g(E_F) = -\frac{g_s g_v e^2}{\pi \hbar v_F}.
\end{equation}
Applying RPA as per Eq.~\eqref{eq:Dyson-chi-2D}, we recover indeed the Luttinger result in Eq.~\eqref{eq:chi-Luttinger}. It is worth noting that the result is independent of the CNT electron density. This can be traced back to the fact that the electron density enters the calculation only through the density of states at the Fermi level, $g(E_F)$, which does not depend on $E_F$ in the linear portion of the Dirac cone (see Fig.~\ref{fig2}b). In Fig.~\ref{fig3}a, we show the tubular response function obtained in the Luttinger/RPA framework. As $q \to 0$, the tubular response function of armchair CNTs converges to 1 (Fig.~\ref{fig3}a). In contrast to the quasi-1D electron gas, the tubular response function of the armchair CNT does not exhibit a cusp at $q=2k_F$, because of the vanishing inter-branch matrix element. Therefore, Friedel oscillations are suppressed in armchair CNTs, at least at the RPA level. Thus, remarkably, armchair CNTs screen the ion potential almost as if they were a perfect metal (Fig.~\ref{fig3}b,c).

We note that the expressions in Eq.~\eqref{eq:Luttinger-charge-Hamiltonian} and Eq.~\eqref{eq:chi-Luttinger} are exact only in the long-wavelength limit $q \to 0$, where the Luttinger Hamiltonian is indeed quadratic; deviations from these results can be expected around $q = 2 k_F$ due to non-Gaussian (power law) correlations\cite{giamarchiQuantumPhysicsOne2003}. Further corrections may arise due to the band dispersion not being exactly linear beyond a certain wavevector ($\abs{q} \approx \SI{0.2}{\angstrom^{-1}}$ for a $(6,6)$ CNT) and due to finite-temperature effects. Nevertheless, these corrections are unlikely to affect our conclusions regarding long-range screening determined by the small-$q$ behaviour. 


\subsection{Rolled-up 2D materials}
The very strong screening behaviour that we obtain in the tubular geometry, even with non-ideal metal models, contrasts with the results obtained in the 2D slit geometry, where the shape of the screened ion potential was found to depend significantly on the confining material's electron density (or Thomas-Fermi length)\cite{kavokineInteractionConfinementElectronic2022}. We now investigate whether this is an effect of geometry, or rather a quantum effect of dimensionality on electronic properties.  

To this end, we introduce a mapping that "rolls up" the density response function of a 2D material $\chi^{\rm 2D}(\vb{q})$ into a tubular geometry with radius $R$:  
\begin{equation}
\label{eq:conversion2D1D}
	\chi^{\rm 1D}_{m,q} = R \chi^{\td}(\vb{q}) \qq{with} \vb{q} = (m/R, q) .
\end{equation}
In this way, we describe a tube that screens Coulomb interactions exactly like its parent 2D material, save for the geometry. 
For a 2D electron gas, the charge density is $n^{\tdeg} = k_F^2 / (2\pi)$, and the static density response function reads \cite{andoElectronicPropertiesTwodimensional1982}
\begin{equation}
\label{eq:chi-2DEG}
	\chi^{\td}(\vb{q}) = -\frac{g_s \me e^2}{2 \pi \hbar^2} \begin{cases}
		1 & \text{, for } \abs{\vb{q}} \leq 2k_F \\
		1 - \sqrt{1 - (2k_F/\abs{\vb{q}})^2} & \text{, for } \abs{\vb{q}} > 2k_F
	\end{cases} .
\end{equation}
For graphene, the charge density is $n^{\rm graph} = k_F^2 / \pi$ and the static density response function reads \cite{hwangDielectricFunctionScreening2007}
\begin{equation}
\label{eq:chi-graphene}
	\chi^{\td}(\vb{q}) = -\frac{g_s \, g_v \, k_F e^2}{2 \pi \hbar v_F} \begin{cases}
		1 & \text{, for } \abs{\vb{q}} \leq 2k_F \\
		1 - \frac{1}{2} \sqrt{1 - \frac{4k_F^2}{q^2}} - \frac{q}{4k_F} \qty( \arcsin(\frac{2k_F}{q}) - \frac{\pi}{2}) & \text{, for } \abs{\vb{q}} > 2k_F
	\end{cases} .
\end{equation}
Fig.~\ref{fig3}a shows the resulting tubular response functions at the same electron density $n$ as the quasi-1D models. Both tubular response functions converge to 1 in the long-wavelength limit $q \to 0$, but at intermediate values of $q$ they are significantly smaller than their 1D counterparts.
Thus, the 2D models screen less effectively than the quasi-1D models (Fig.~\ref{fig3}b,c). We conclude that the predicted qualitative difference in screening between the 1D tube and 2D slit settings is not only a question of geometry: the quantum confinement of electrons along the tube circumference -- without which the mapping in Eq.~\eqref{eq:conversion2D1D} would be exact -- is instrumental in the nearly perfect metallic behaviour of armchair CNTs.

\section{Self-energy}
Until now, we have studied the screened ion potential as a function of distance from the ion along the tube axis. However, an ion also experiences a change in its electrostatic self-energy upon entering a polarisable nanotube, due to two effects. First, the dielectric background provided by water changes from isotropic bulk water with $\epsb \approx 80$ to anisotropic nanoconfined water, for which we assume $\epsz = 80$ and $\epsp = \epsr = 2$, as in previous sections. We estimate the resulting self-energy by computing the difference in Born energy between the two backgrounds, $\Sigma_{\epsilon} = \Sigma_{\text{1D}} - \Sigma_{\text{bulk}}$. Second, the induced potential of the solid acts back on the ion itself according to Eq.~\eqref{eq:induced-ion-potential}, i.e. $\Sigma_{\text{NT}} = \phi^{\ind}(\vb{r}=\vb{0})$.
We model the ion as a homogeneously charged ball with radius $R_0$ and total charge $Qe$. Its charge density is thus $n = eQ/V = 3eQ/(4\pi R_0^3)$.

In an anisotropic dielectric background, whose dielectric tensor $\vb*{\epsilon}$ is diagonal with entries $\epsz \neq \epsr = \epsp$, the Laplace equation for the electrostatic potential $\phi$ reads $\nabla [\vb*{\epsilon} \nabla \phi(\vb{r})] = 0$. Rescaling the space coordinates into $x' = x$, $y' = y$ and $z' = \alpha z$ with $\alpha = \sqrt{\epsr / \epsz}$ yields an isotropic dielectric constant $\epsr$. However, it deforms the spherical ion into an ellipsoid with half-axes $a=b=R_0$ and $c=\alpha R_0$, and with charge density $n' = eQ/V' = n / \alpha$, where $V'=\frac{4}{3} \pi \, abc$ denotes the volume of the ellipsoid. The self-energy is defined as the charge density of the ion interacting with its own potential $\phi(\vb{r})$, 
\begin{equation}
\label{eq:self-energy-definition}
	\Sigma_{\text{1D}} = \frac{1}{2} \int_{V'} n' \, \phi(\vb{r'}) \dd[3]{r'},
\end{equation}
The electrical potential of a homogeneously charged ellipsoid inside its interior is \cite{chandrasekharEllipsoidalFiguresEquilibrium1969}
\begin{equation}
\label{eq:potential-ellipsoid}
	\phi(\vb{r'}) = \frac{1}{4 \epsilon_0 \epsilon_r} abc \, n' \int_0^{\infty} \dd{s}  \frac{1 - \frac{x'^2}{a^2 + s} - \frac{y'^2}{b^2 + s} - \frac{z'^2}{c^2 + s}}{\sqrt{(a^2 + s) (b^2 + s) (c^2 + s)} }.
\end{equation}
Thus, the integral in Eq.~\eqref{eq:self-energy-definition} can be evaluated using the formula for the second moments of an ellipsoid. This yields
\begin{equation}
\label{eq:self-energy-confined}
	\Sigma_{\text{1D}} = \frac{3}{5} \frac{(Qe)^2}{4\pi\epsilon_0\epsilon_r R_0} \frac{\arccos(\sqrt{\epsr / \epsz})}{\sqrt{1 - \epsr / \epsz}}.
\end{equation}
The self-energy in a homogeneous and isotropic dielectric background with dielectric constant $\epsilon$ is known to be
\begin{equation}
\label{eq:self-energy-bulk}
	\Sigma_{\text{bulk}} = \frac{3}{5} \frac{(Qe)^2}{4\pi \epsilon_0 \epsb R_0},
\end{equation}
which coincides with Eq.~\eqref{eq:self-energy-confined} for $\epsr \to \epsz = \epsb$.

\begin{figure}
	\centering
	\includegraphics{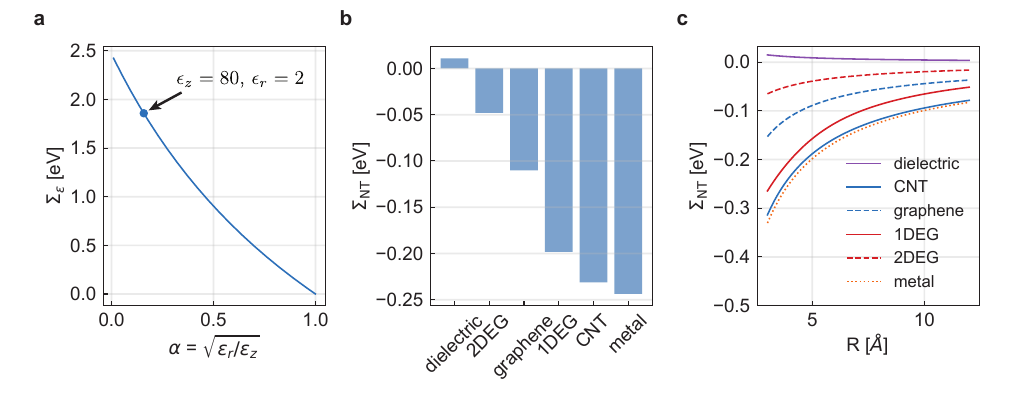}
	\caption{\textbf{Self-energy of a unit-charge ion with radius $R_0=\SI{1}{\angstrom}$ at the centre of a polarisable nanotube}. \textbf{a} Self-energy $\Sigma_{\epsilon}$ due to the change in dielectric background between bulk and nanoconfined water, as a function of the anisotropy factor $\alpha = \sqrt{\epsr/\epsz}$ assuming $\epsz = \epsilon_{\text{bulk}}$. The blue dot corresponds to the choice of dielectric constants used in this paper, $\epsz=80$ and $\epsr=2$. \textbf{b} Nanotube polarisation contribution $\Sigma_{\text{NT}}$ to the self-energy of an ion at the centre of various nanotubes with radius $R=\SI{4}{\angstrom}$ and identical parameters as in Fig.~\ref{fig3}. \textbf{c} Same as \textbf{b}, here depending on the tube radius $R$.
	}
	\label{fig4}
\end{figure}

Fig.~\ref{fig4}a shows the water contribution to the self-energy $\Sigma_{\epsilon} = \Sigma_{\text{1D}} - \Sigma_{\text{bulk}}$ for a monovalent ion ($Q=1$) with radius $R_0 = \SI{1}{\angstrom}$ versus the anisotropy factor $\alpha = \sqrt{\epsr/\epsz}$. 
In \aa ngstr\"om-scale nanotubes, there is strong anisotropy\cite{locheGiantAxialDielectric2019}, corresponding to small values of $\alpha$ and a large self-energy. Upon increasing the tube radius to $R \sim \SI{10}{\nm}$, anisotropy gradually disappears and $\alpha$ converges to its bulk value $\alpha=1$, leading to $\Sigma_{\epsilon} = 0$.
Fig.~\ref{fig4}b shows the nanotube screening contribution $\Sigma_{\text{NT}}$, for all of our previous nanotube models, with radius $R \approx \SI{4}{\angstrom}$. This contribution can be either positive or negative, depending on the electronic properties of the nanotube \cite{kavokineInteractionConfinementElectronic2022}. Metallic walls yield negative self-energies, thereby reducing the energy cost of an ion entering the nanotube. In contrast, a purely dielectric tube wall can give a positive contribution to the self-energy, further repelling the ion from the nanotube. In both cases, the absolute value decreases with increasing tube radius, as the centrally-placed ion is further away from the wall (Fig~\ref{fig4}c). With the parameters used throughout the paper, $R \approx \SI{4}{\angstrom}$, $R_0 = \SI{1}{\angstrom}$ and $\alpha \approx 0.16$, the total self-energy is dominated by the strongly anisotropic dielectric background of water, to which the nanotube polarisability only provides a small correction.

\section{Conclusion}
In this paper, we have demonstrated that the presence of a polarisable nanotube can fundamentally alter Coulomb interactions in quasi-1D confinement. Due to the substantial anisotropy of the dielectric constant of water in \aa ngstr\"om-scale nanotubes, Coulomb interactions are strongly enhanced. A dielectric nanopore material with a low dielectric constant further enhances Coulomb interactions, whereas a metallic nanotube material strongly screens the long-range part of the Coulomb potential. We applied Luttinger liquid theory to show that this behaviour persists under the exact treatment of electron-electron interactions in realistic 1D nanotubes: at intermediate distances, we found that a quasi-1D electron gas and a single-walled carbon nanotube screen the Coulomb potential essentially like a perfect metal.
This perfect screening property cannot be reduced to a geometric effect and relies on the quantum confinement of electrons along the tube circumference. At longer distances, the quasi-1D electron gas displays Friedel oscillations, while these are suppressed in a metallic armchair CNT due to the sublattice structure.
Accounting for the substantial anisotropy of water in quasi-1D confinement, we found a large self-energy barrier for an ion to enter the nanotube, which is only weakly modified by the dielectric properties of the nanotube walls. 

Our approach is based on tubular response functions, which generalise surface response functions to the nanotube, arguably the most widespread geometry in nanofluidics. Our framework can be applied to compute Coulomb interactions in realistic nanotube models, including carbon and boron nitride nanotubes. While the case of a metallic carbon nanotube has been explicitly presented, the 1D-electron gas can serve as a simple model for highly doped nanotubes, where the Fermi level lies within bands with approximately quadratic dispersion. We also expect our findings to qualitatively extend to nanoporous materials, such as carbide-derived carbon, a promising candidate for supercapacitors \cite{chmiolaMonolithicCarbideDerivedCarbon2010}. For large tube radii, our theory converges to the theory of surface response functions for planar interfaces \cite{kavokineInteractionConfinementElectronic2022}.

A substantial simplification in our theory is treating water as a dielectric background. This macroscopic model is known to break down in the smallest nanotubes, where the water molecules form a single 1D chain \cite{locheGiantAxialDielectric2019}.
MD-simulations of nano-confined water show that the axial dielectric constant oscillates close to the wall due to layering effects, similar to planar interfaces, and also the radial dielectric constant has a non-monotonous radius-dependence \cite{locheGiantAxialDielectric2019}. However, even for the smallest possible tubes, uniform effective dielectric constants have been extracted from these simulations and can be used as an input to our theory \cite{locheGiantAxialDielectric2019}. Short-range electrostatic interactions, in particular the ion self-energy, could be sensitive to the local molecular configuration of the water in the presence of the ion, which is not captured by effective dielectric constants. However, we expect long-range interactions to be largely unaffected, as they are dominated by the polarisability of the solid. 
More generally, our framework is not restricted to ions in a homogeneous background. It can be applied to any electrolyte or polar liquid whose molecules interact via Coulombic forces, such as water itself.
Such a treatment could test the assumption of a homogeneous dielectric background and will be left for future investigations.

In this paper, we used the tubular response function of the solid as an input. We also derived a tubular response function of the liquid. In principle, the framework could be generalised to the case in which the solid and the liquid mutually renormalise each other's responses. A dynamical treatment of this effect is closely related to the fluctuation-induced quantum friction force \cite{kavokineFluctuationinducedQuantumFriction2022}. Thus, our framework paves the way for quantitative predictions of electrolyte structure and dynamics in one-dimensional nanofluidic systems.

\section*{Data availability}
All data associated with this study is available on Zenodo: \href{https://doi.org/10.5281/zenodo.20128754}{https://doi.org/10.5281/zenodo.20128754}. 

\section*{Author Contributions}
Conceptualisation: Nikita Kavokine and Peter Gispert.
Methodology: Peter Gispert and Nikita Kavokine.
Investigation: Peter Gispert.
Writing -- Original Draft: Peter Gispert.
Writing -- Review \& Editing: Peter Gispert and Nikita Kavokine.
Supervision: Nikita Kavokine.

\section*{Conflicts of interest}
There are no conflicts to declare.







\bibliography{Ion-in-NT.bib} 

\bibliographystyle{rsc} 

\end{document}